\documentclass[aps,pra,preprint,superscriptaddress]{revtex4}
\usepackage{amsmath}
\usepackage{amssymb}
\usepackage{graphicx}
\usepackage{dcolumn}
\usepackage{bm}

\mathchardef\ogon="012C
\newcommand{\as}{a\kern-0.22em\lower.40ex\hbox{$_{\ogon}$}}

\begin{document}

\title{Density-density correlation and interference mechanism for two
initially independent Bose-Einstein condensates}
\author{Shujuan Liu}
\affiliation{State Key Laboratory of Magnetic Resonance and Atomic and Molecular Physics,
Wuhan Institute of Physics and Mathematics, Chinese Academy of Sciences,
Wuhan 430071, P. R. China}
\affiliation{Center for Cold Atom Physics, Chinese Academy of Sciences, Wuhan 430071, P.
R. China }
\affiliation{Graduate School of the Chinese Academy of Sciences, P. R. China}
\date{\today }
\author{Hongwei Xiong\footnote{
xionghongwei@wipm.ac.cn}}
\affiliation{State Key Laboratory of Magnetic Resonance and Atomic and Molecular Physics,
Wuhan Institute of Physics and Mathematics, Chinese Academy of Sciences,
Wuhan 430071, P. R. China}
\affiliation{Center for Cold Atom Physics, Chinese Academy of Sciences, Wuhan 430071, P.
R. China }

\begin{abstract}
In an exciting experiment by MIT's group (Science \textbf{275}, 637 (1997)),
clear interference fringes were observed for two initially independent Bose
condensates in dilute gas. Presently, there are two different theories
(measurement-induced interference theory and interaction-induced
interference theory) which can both explain MIT's experimental results.
Based on our interaction-induced interference theory, we consider the
evolution of the density-density correlation after the releasing of a
double-well potential trapping two independent Bose condensates. Based on
the interaction-induced interference theory, we find that the interference
fringes in the density-density correlation exhibit a behavior of emergence
and disappearance with the development of time. We find essential difference
for the density-density correlation based on interaction-induced
interference theory and measurement-induced interference theory, and thus we
suggest the density-density correlation to experimentally reveal further the
interference mechanism for two initially independent Bose condensates.

\end{abstract}

\maketitle

\section{introduction}

The interference between two initially spatially-separated Bose-Einstein
condensates (BECs) is a fundamental physical problem \cite{Andrew}. The
potential application of atom interferometry makes the theoretical and
experimental researches on the interference mechanism become quite
attractive. On the other hand, the theoretical and experimental researches
in the last ten years show that the physical mechanism of the interference
phenomena between two initially independent Bose condensates is a quite
nontrivial problem. In fact, there exist considerable controversies about
the physical mechanism of the interference phenomena for two initially
independent Bose condensates. Presently, a popular interpretation about the
interference patterns observed in \cite{Andrew} for two initially
independent condensates is the so-called measurement-induced interference
theory \cite{JAV,Zoller,CASTIN,Leggett,Pethick,Stringari}. Recently, we
proposed a new interpretation of interaction-induced interference theory
\cite{xiong1,xiong2,xiong-NJP,xiong-coh}, which agrees well with the MIT's
experimental results \cite{Andrew}. Recently, L. S. Cederbaum \textit{et al}
\cite{Band} supported the viewpoint that interatomic interaction plays an
essential role in the observed interference phenomena. Most recently, the
theoretical work by D. Masiello and W. P. Reinhardt \cite{Reinhardt} also
showed the emergence of interference fringes in the presence of interatomic
interaction.

Considering the fact that the interference mechanism between two
initially independent Bose condensates is a fundamental and subtle
question, and the fact that two completely different theories (i.e.
measurement-induced interference theory and interaction-induced
theory) coexist, an important and urgent problem is to propose an
experimental suggestion to test two different theories. In this
paper, we give an experimental scheme to test two theories in future
experiment based on the calculations of the density-density
correlation. Different from our previous works about the
interaction-induced interference theory \cite%
{xiong1,xiong2,xiong-NJP,xiong-coh} and other works supporting this theory
\cite{Band,Reinhardt}, to the best of our knowledge, the present paper gives
the first calculations about the density-density correlation in the frame of
interaction-induced interference theory.

In MIT's experiment \cite{Andrew}, the density distribution (first-order
coherence) was measured after the overlapping between two initially
independent Bose condensates. In the last few years, intensive theoretical
\cite{Glauber,Lukin,Strin,lattice-theo,Fin-tem,dynamics-1D,REV} and
experimental studies \cite%
{Shimizu,coherence-bec,HBT-Cold,Esslinger,Jin,phase-fluc,Elonged,Bloch,Phillips,Bloch-fermi,aspect-fermion}
have shown that high-order correlation is very powerful in revealing the
many-body quantum characteristics of ultracold atomic system. For example,
the second-order correlation function was studied experimentally in \cite%
{Esslinger} to reveal whether atom lasers exhibit a truly laserlike
behavior. We believe it is worthwhile to study the high-order correlation
for two initially independent Bose condensates, and (up to our knowledge)
the previous theoretical studies about the role of interatomic interaction
in the interference mechanism did not address this sort of problem yet \cite%
{xiong1,xiong2,xiong-NJP,xiong-coh,Band,Reinhardt}. Based on the
interaction-induced interference theory, we calculate the evolution of the
density-density correlation after releasing a double-well potential trapping
two independent condensates. We find that the interference fringes in the
density-density correlation exhibit a special behavior of emergence and
disappearance with the development of time. Besides this interesting
behavior, we also notice that there is an essential difference in the
density-density correlation between interaction-induced interference theory
and measurement-induced interference theory, and thus we suggest the
experimental studies of the density-density correlation to test further the
interference mechanism of two initially independent Bose condensates.

Because the interference mechanism is a quite subtle problem, we think it
would be quite useful to give a brief introduction to the problem,
experimental facts and different theories. About ten years ago, the MIT
group gave strong evidence about the spatial coherence for Bose condensates
through the observation of clear interference patterns for two separated
condensates \cite{Andrew}. In MIT's experiment, two separated condensates
were prepared in a double-well system after evaporative cooling. After
releasing the double-well trapping potential, two condensates expand and
overlap with sufficiently long expanding time. After their overlapping, the
MIT group observed clear interference patterns for two completely different
situations. (i) When the chemical potential of the system is larger than the
height of the center barrier of the double-well potential, the two
condensates prepared after the evaporative cooling can be regarded to be
coherently separated. In this situation, every atom is described by an
identical wave function which is a coherent superposition of two wave
packets in different wells. It is not surprising that clear interference
patterns can be observed. In fact, the MIT group did observe clear
interference patterns in this situation. (ii) If the chemical potential of
the system is much smaller than the height of the center barrier so that the
tunneling between two wells can be omitted, after the evaporative cooling
for the cold atomic cloud in the double-well trap, two condensates prepared
are spatially-separated and independent. Here \textquotedblleft
independent\textquotedblright\ means that any atom either exists in the left
well, or exists in the right well. It is understandable that no atom's
quantum state is a coherent superposition of the wave functions in two
separated wells, because during the evaporative cooling precess, the losses
of the atoms and interatomic collisions lead to strong decoherence between
two wells, and the negligible tunneling can not restore the coherence
between two wells even at zero temperature. In this situation, the initial
quantum state of the whole system at zero temperature is widely described by
a Fock state $\left\vert N_{1},N_{2}\right\rangle $, with $N_{1}$ and $N_{2}$
being the number of particles in the left and right wells. The most exciting
phenomenon in MIT's experiment is the observation of clear interference
patterns for the later situation (two initially independent condensates).
For two initially independent condensates, before the releasing of the
double-well potential, there is no coherence between two condensates.
Exaggeratedly speaking, this is similar to the situation that there is no
coherence or correlation between two condensates separately prepared in two
different experimental apparatuses. For each condensate, there are no
interference patterns during the free expansion. The interference patterns
emerge after the overlapping between two condensates. Thus, the experimental
observation of the interference patterns means that a coherence between two
initially independent condensates must be established by a physical
mechanism. The fundamental question is what's the physical mechanism for
this coherence-establishment process.

Presently, the most popular viewpoint is the so-called measurement-induced
interference theory \cite{JAV,Zoller,CASTIN,Leggett,Pethick,Stringari}.
Measurement-induced interference theory thinks that before the measurement
of the density distribution, there is no interference term in the density
expectation value even after the overlapping between two initially
independent condensates. In measurement-induced interference interpretation,
it is thought that the measurement process and the interference terms in the
two-particle correlation $\left\langle N_{1},N_{2}\right\vert \widehat{\Psi }%
^{\dag }(\mathbf{r})\widehat{\Psi }^{\dag }(\mathbf{r}^{\prime })\widehat{%
\Psi }(\mathbf{r}^{\prime })\widehat{\Psi }(\mathbf{r})\left\vert
N_{1},N_{2}\right\rangle $ would establish the coherence between two
overlapping and initially independent condensates. When the number of
particles is much larger than $1$, with more and more particles being
detected, the measurement induces the transformation from the Fock state to
a phase state: $\left\vert N/2,N/2\right\rangle \Longrightarrow \left(
\widehat{a}_{1}^{\dag }e^{i\varphi /2}+\widehat{a}_{2}^{\dag }e^{-i\varphi
/2}\right) ^{N}\left\vert 0\right\rangle /\left( 2^{N}N!\right) ^{1/2}$.
Here $\widehat{a}_{1}^{\dag }$ and $\widehat{a}_{2}^{\dag }$ are the
creation operators for the left and right condensates, and $\varphi $ is a
random phase factor. After this transformation, it is argued that there
would be interference patterns in the measurement result of the density
distribution. In measurement-induced interference interpretation, the
interference terms in two-particle correlation function play a key role in
the formation process of the phase state (a clear introduction to this
process can be found in \cite{Pethick}).

Recently, we found that interatomic interaction can establish the coherence
between two initially independent condensates after their overlapping \cite%
{xiong1,xiong2,xiong-NJP,xiong-coh}.\textit{\ }In our interaction-induced
interference theory \cite{xiong1,xiong2,xiong-NJP}, it is argued that when
interatomic interaction is considered, with the evolution of the whole
quantum state (the evolution of the quantum state is obtained by solving the
many-body Schr\H{o}dinger equation), the interference patterns emerge
definitely in the density expectation value, after the overlapping between
two initially independent condensates. This means that the relative phase
(note that this relative phase is completely different from the
insignificant relative phase between two initially independent condensates
before the releasing of the double-well potential) between two condensates
after the establishment of the coherence is definite, and can be known in
principle when the initial condition is known. In the measurement-induced
interference theory, however, the relative phase $\varphi $ is random in
every single-shot experiment. Thus, observing whether there is random
relative phase (i.e. random shift of the interference patterns) in different
experiments can give us direct test between measurement-induced interference
theory and interaction-induced interference theory. However, more careful
analyses show that this sort of experiment may be quite challenging. Even
for completely identical initial condition for the cold atomic cloud in the
double-well trap, after the evaporative cooling, there are unavoidable
particle-number fluctuations for the condensates in two wells (i.e. for
every experiment, the particle number in each well is different). Roughly
speaking, if the interaction-induced interference theory is correct, the
particle-number fluctuations will lead to a random relative phase
approximated as $\left\vert \Delta \mu t/\hbar \right\vert $ ($\Delta \mu $
is the difference of the chemical potential for two condensates). If $%
\left\vert \Delta \mu t/\hbar \right\vert \gtrsim \pi $, the experimental
observation of the random shift of interference patterns could not give
definite evidence to distinguish the measurement-induced interference theory
and interaction-induced interference theory. For the parameters in MIT's
experiment, simple estimation shows that for the expansion time of $t_{\exp
}=40$ \textrm{ms}, $\left\vert \Delta \mu t/\hbar \right\vert \approx 20\pi
\left\vert \triangle N/N\right\vert $ \cite{note1}. Here $\triangle N/N$ is
the relative difference of the particle number in two wells. In MIT's
experiment, rough estimation shows that $\left\vert \triangle N/N\right\vert
$ is larger than $2\%$ \cite{note2}. In this situation, even random shift of
the interference patterns was observed in MIT's experiment, the
interaction-induced interference theory can not be excluded. Other technical
noises would further challenge the test of two theories based on the
measurement of the random relative phase in the density distribution. M. A.
Kasevich \cite{Kasevich} gave a very interesting discussion about the
technical noise and random phase shift by pointing out that `In the
experiments of \cite{Andrew}, the phase of the observed interference pattern
did fluctuate, though it was not possible to unambiguously identify the
mechanism of the fluctuations, because technical noise sources were also
present in the measurements.'.

The above analyses show that further experimental studies are needed to
reveal the interference mechanism for two initially independent condensates.
In measurement-induced interference theory, when $N_{1}\gg 1$ and $N_{2}\gg
1 $, there would be clear interference fringes after the overlapping of two
initially independent condensates. In our interaction-induced interference
theory, even for the situation of $N_{1}\gg 1$ and $N_{2}\gg 1$, there would
be no observable interference fringes for sufficiently small interaction
effect. In the extreme noninteracting situation, the interaction-induced
interference theory gives a prediction that there is always no interference
fringes in the density distribution for the initial Fork state $\left\vert
N_{1},N_{2}\right\rangle $. In an experiment, by adjusting the particle
number, separation between two wells and even interatomic interaction
through Feshbach resonance, one may test two different theories from the
measurement of the density distribution.

Considering the fact that the interference mechanism is a very subtle
problem, besides the test based on first-order spatial coherence, we believe
high-order coherence can provide further important and at least
complementary information about the interference mechanism. This is a little
similar to the discrimination between a thermal light (a series of
independent photons) and a laser beam based on the high-order correlation.
Another merit of the density-density correlation lies in that for the
experimental conditions and parameters in the original work of Ref. \cite%
{Andrew}, we can give important test about the interaction-induced
interference theory and measurement-induced interference theory based on the
experimental studies of the density-density correlation.

We consider the following normalized density-density correlation function%
\begin{equation}
g_{nn}(\mathbf{d},t)=\frac{\int \left\langle \widehat{n}(\mathbf{r}+\mathbf{d%
}/2,t)\widehat{n}(\mathbf{r}-\mathbf{d}/2,t)\right\rangle dV}{\int
\left\langle \widehat{n}(\mathbf{r}+\mathbf{d}/2,t)\right\rangle
\left\langle \widehat{n}(\mathbf{r}-\mathbf{d}/2,t)\right\rangle dV}.
\label{density-density correlation}
\end{equation}%
This sort of density-density correlation was studied recently for ultracold
bosons \cite{Bloch,Phillips}\textit{\ }and fermions \cite{Bloch-fermi}
released from an optical lattice. Based on the interaction-induced
interference theory, at the beginning of the overlapping and due to the
presence of interference term in the density-density correlation $%
\left\langle \widehat{n}(\mathbf{r}+\mathbf{d}/2,t)\widehat{n}(\mathbf{r}-%
\mathbf{d}/2,t)\right\rangle $, there would be interference structure in $%
g_{nn}(\mathbf{d},t)$. With further time evolution and when two condensates
become fully coherent, the quantum state can be approximated as $\left\vert
N\right\rangle $. $g_{nn}(\mathbf{d},t)$ would become flat in this
situation. This means that there is a behavior of emergence and
disappearance for the interference fringes in $g_{nn}(\mathbf{d},t)$ based
on the interaction-induced interference theory. In the measurement-induced
interference theory, however, after the overlapping between two condensates,
$g_{nn}(\mathbf{d},t)$ would always exhibit a flat behavior because the
measurement makes the system become a phase state. These simple analyses
show clearly that there is essential difference in $g_{nn}(\mathbf{d},t)$
between measurement-induced interference theory and interaction-induced
interference theory. A merit in the measurement of the density-density
correlation function is that small particle-number fluctuations during the
evaporative cooling will not change significantly the structure of the
density-density correlation function.

The paper is organized as follows. In Sec. II, we give the result of
density-density correlation function based on the interaction-induced
interference theory, while in Sec. III we give the prediction of
density-density correlation function based on measurement-induced
interference theory. In the last section, a brief summary and discussion are
given.

\section{density-density correlation function in interaction-induced
interference theory}

Coherence establishment mechanism due to interparticle interaction is in
fact widely known and appears in a large number of physical phenomena. A
famous example is the quantum phase transition from Mott insulator state to
superfluid state for ultracold atoms in optical lattices \cite%
{Bose-Hubb,Jaksch,QPT-exp}, when the strength of the optical lattices is
decreased adiabatically below a critical value. One should note that
interatomic interaction plays key role in this quantum phase transition
process. Without the interatomic interaction, decreasing the strength of the
optical lattices can make the wavepacket of the atoms exist\ in the whole
optical lattices, but all the atoms are in different and orthogonal quantum
states. For the double-well system, the adiabatic varying of the center
barrier has been studied intensively \cite{Milburn,Spekk,Javan,Menotti,Yi,Ho}%
. These theoretical studies show clearly that interparticle interaction and
the adiabatic decreasing of the center barrier make the initial Fock state $%
\left\vert N/2,N/2\right\rangle $ become a coherently superposed state $%
\left\vert N\right\rangle $ where the quantum state of all atoms become a
coherent superposition of two wavepackets in different wells. Most recently,
we considered the general situation with unequal particle number in two
wells, and the interaction-induced coherence process was also found,
especially the Josephson effect for too initially independent Bose
condensates was predicted \cite{xiong-coh}. In MIT's experiment, the
double-well potential for two initially independent condensates was switched
off suddenly to observe the interference patterns. This releasing of the
double-well potential can be regarded as a non-adiabatical decreasing of the
center barrier. Although non-adiabatic process can be quite different from
an adiabatic process, the interaction-induced coherence mechanism for the
adiabatic process of the double-well system gives us inspiration that
interatomic interaction could play important role in the interference
phenomena observed in MIT's experiment.

In our interaction-induced interference theory \cite{xiong1,xiong2,xiong-NJP}%
, the interatomic interaction plays a role of establishing the spatial
coherence between two initially independent condensates. We assume that
initially there are $N_{1}$ particles in the left condensate and $N_{2}$
particles in the right condensate. We assume further that the
single-particle wave functions are respectively $\phi _{1}$ and $\phi _{2}$.
After releasing the double-well potential trapping two independent
condensates, even after their overlapping, $\phi _{1}$ and $\phi _{2}$ are
always orthogonal without interatomic interaction. In the presence of
interatomic interaction, however, $\phi _{1}$ and $\phi _{2}$ are no longer
orthogonal after their overlapping. If we still use two orthogonal basis $%
\phi _{1}$ and $\phi _{2}^{\prime }$, the non-orthogonal property between $%
\phi _{1}$ and $\phi _{2}$ is equivalent to the physical picture that there
are coherent particle exchanges between the orthogonal basis $\phi _{1}$ and
$\phi _{2}^{\prime }$ in the presence of interatomic interaction. We have
verified rigorously that for large particle number, even a small
non-orthogonality between $\phi _{1}$ and $\phi _{2}$ can lead to
high--contrast interference fringes in the density expectation value $%
\left\langle \widehat{n}(\mathbf{r},t)\right\rangle $ \cite%
{xiong1,xiong2,xiong-NJP}.

When the non-orthogonal property between $\phi _{1}$ and $\phi _{2}$ is
considered, it is obvious that $\widehat{a}_{1}=\int \phi _{1}^{\ast }%
\widehat{\Psi }dV$ and $\widehat{a}_{2}^{\dag }=\int \phi _{2}\widehat{\Psi }%
^{\dag }dV$ are not commutative any more. Simple calculation gives $[%
\widehat{a}_{1},\widehat{a}_{2}^{\dagger }]=\zeta ^{\ast }=\int \phi
_{1}^{\ast }\phi _{2}dV$. In this situation, the many-body quantum state is%
\begin{equation}
\left\vert N_{1},N_{2}\right\rangle =\frac{\Xi _{n}}{\sqrt{N_{1}!N_{2}!}}(%
\widehat{a}_{1}^{\dag })^{N_{1}}(\widehat{a}_{2}^{\dag })^{N_{2}}\left\vert
0\right\rangle  \label{quantum state sum}
\end{equation}
$\Xi _{n}$ is a normalization constant determined by%
\begin{equation}
\Xi _{n}^{2}\left( \sum\limits_{i=0}^{N_{2}}\frac{N_{2}!\left(
N_{1}+i\right) !\left( 1-\left\vert \zeta \right\vert ^{2}\right)
^{N_{2}-i}\left\vert \zeta \right\vert ^{2i}}{i!i!N_{1}!\left(
N_{2}-i\right) !}\right) =1.
\end{equation}%
It is well-known that the field operator should be expanded in terms of a
complete and orthogonal basis set. Therefore, we construct two orthogonal
wave functions $\phi _{1}$ and $\phi _{2}^{\prime }$. Assuming that $\phi
_{2}^{\prime }=\beta \left( \phi _{2}+\alpha \phi _{1}\right) $, based on
the conditions $\int \phi _{1}^{\ast }\phi _{2}^{\prime }dV=0$ and $\int
\left\vert \phi _{2}^{\prime }\right\vert ^{2}dV=1$, we have $\left\vert
\beta \right\vert =\left( 1-\left\vert \zeta \right\vert ^{2}\right) ^{-1/2}$
and $\alpha =-$ $\zeta ^{\ast }$. Introducing an annihilation operator $%
\widehat{k}=\int \widehat{\Psi }\left( \phi _{2}^{\prime }\right) ^{\ast }dV$%
, $\widehat{a}_{1}$ and $\widehat{k}^{\dag }$ are commutative. In this
situation, the field operator is expanded as%
\begin{equation}
\widehat{\Psi }=\widehat{a}_{1}\phi _{1}+\widehat{k}\phi _{2}^{\prime
}+\cdots .  \label{new-expansion}
\end{equation}

The evolution equation is obtained by first obtaining the energy expression,
and then using the action principle. The overall energy is%
\begin{equation}
E=\left\langle N_{1},N_{2},t\right\vert \widehat{H}\left\vert
N_{1},N_{2},t\right\rangle .
\end{equation}%
Here $\left\vert N_{1},N_{2},t\right\rangle $ is given by Eq. (\ref{quantum
state sum}), while $\widehat{H}$ is the Hamiltonian of the system which
takes the following form%
\begin{equation}
\widehat{H}=\int dV\left( \frac{\hbar ^{2}}{2m}\nabla \widehat{\Psi }^{\dag
}\cdot \nabla \widehat{\Psi }+\frac{g}{2}\widehat{\Psi }^{\dag }\widehat{%
\Psi }^{\dag }\widehat{\Psi }\widehat{\Psi }\right) .
\end{equation}%
By using the ordinary action principle and the energy of the whole system,
we have the following coupled evolution equations for $\phi _{1}$ and $\phi
_{2}$ \cite{xiong-NJP}:
\begin{eqnarray}
i\hslash \frac{\partial \phi _{1}}{\partial t} &=&\frac{1}{N_{1}}\frac{%
\delta E}{\delta \phi _{1}^{\ast }},  \label{GP1} \\
i\hslash \frac{\partial \phi _{2}}{\partial t} &=&\frac{1}{N_{2}}\frac{%
\delta E}{\delta \phi _{2}^{\ast }},  \label{GP2}
\end{eqnarray}%
where $\delta E/\delta \phi _{1}^{\ast }$ and $\delta E/\delta \phi
_{2}^{\ast }$ are functional derivatives.

When the time evolution of $\phi _{1}$ and $\phi _{2}$ is obtained, the
evolution of the density distribution is then
\begin{eqnarray}
n\left( \mathbf{r},t\right) &=&\left\langle N_{1},N_{2},t\right\vert
\widehat{n}\left\vert N_{1},N_{2},t\right\rangle  \notag \\
&=&\alpha _{d}\left\vert \phi _{1}\left( \mathbf{r},t\right) \right\vert
^{2}+2\beta _{d}\times \mathrm{Re}\left( e^{i\varphi _{c}}\phi _{1}^{\ast
}\left( \mathbf{r},t\right) \phi _{2}\left( \mathbf{r},t\right) \right)
+\gamma _{d}\left\vert \phi _{2}\left( \mathbf{r},t\right) \right\vert ^{2},
\label{ndensity}
\end{eqnarray}%
where the coefficients are
\begin{eqnarray}
\alpha _{d} &=&\sum\limits_{i=0}^{N_{2}}\frac{\Xi _{n}^{2}N_{2}!\left(
N_{1}+i-1\right) !N_{1}\left( 1-\left\vert \zeta \right\vert ^{2}\right)
^{N_{2}-i}\left\vert \zeta \right\vert ^{2i}}{i!i!\left( N_{1}-1\right)
!\left( N_{2}-i\right) !},  \label{adS} \\
\beta _{d} &=&\sum\limits_{i=0}^{N_{2}-1}\frac{\Xi _{n}^{2}N_{2}!\left(
N_{1}+i\right) !\left( 1-\left\vert \zeta \right\vert ^{2}\right)
^{N_{2}-i-1}\left\vert \zeta \right\vert ^{2i+1}}{i!\left( i+1\right)
!\left( N_{1}-1\right) !\left( N_{2}-i-1\right) !},  \label{bdS} \\
\gamma _{d} &=&\sum\limits_{i=0}^{N_{2}-1}\frac{\Xi _{n}^{2}N_{2}!\left(
N_{1}+i\right) !\left( 1-\left\vert \zeta \right\vert ^{2}\right)
^{N_{2}-i-1}\left\vert \zeta \right\vert ^{2i}}{i!i!N_{1}!\left(
N_{2}-i-1\right) !}.  \label{cdS}
\end{eqnarray}%
Here, the relative phase $\varphi _{c}$ is determined by $e^{i\varphi
_{c}}=\zeta /\left\vert \zeta \right\vert $. It is clearly shown that for
nonzero $\zeta $, there is a nonzero interference pattern in $n\left(
\mathbf{r},t\right) $. For $N_{1}\left\vert \zeta \right\vert \gg 1$ and $%
N_{2}\left\vert \zeta \right\vert \gg 1$, we have proven that $n\left(
\mathbf{r},t\right) $ can be approximated very well as \cite%
{xiong1,xiong2,xiong-NJP}%
\begin{equation}
n\left( \mathbf{r},t\right) \approx \left\vert \sqrt{N_{1}}\phi _{1}\left(
\mathbf{r},t\right) +\sqrt{N_{2}}e^{i\varphi _{c}}\phi _{2}\left( \mathbf{r}%
,t\right) \right\vert ^{2}.  \label{density-coherent}
\end{equation}%
In this situation, $\Phi _{e}=\sqrt{N_{1}}\phi _{1}+\sqrt{N_{2}}e^{i\varphi
_{c}}\phi _{2}$ can be regarded an effective order parameter of the whole
system. For $N_{1}\left\vert \zeta \right\vert \gg 1$ and $N_{2}\left\vert
\zeta \right\vert \gg 1$, we have also proven that $\Phi _{e}$ satisfies the
ordinary Gross-Pitaevskii equation (a detailed analysis can be found in \cite%
{xiong-NJP})%
\begin{equation}
i\hslash \frac{\partial \Phi _{e}}{\partial t}\simeq -\frac{\hslash ^{2}}{2m}%
\nabla ^{2}\Phi _{e}+g\left\vert \Phi _{e}\right\vert ^{2}\Phi _{e}.
\label{S-app-equation}
\end{equation}%
In brief, in interaction-induced interference theory, the interatomic
interaction establishes the coherence between two initially independent
condensates, and thus leads to the emergence of the nonzero interference
term in the density expectation value. The interference fringes based on our
interaction-induced interference theory agree well with the MIT's experiment
\cite{xiong1,xiong-NJP}.

After straightforward calculations, the normalized density-density
correlation function is given by%
\begin{equation}
g_{nn}\left( \mathbf{d},t\right) =\frac{\int \left\langle
N_{1},N_{2},t\right\vert \widehat{n}(\mathbf{r}+\mathbf{d}/2,t)\widehat{n}(%
\mathbf{r}-\mathbf{d}/2,t)\left\vert N_{1},N_{2},t\right\rangle dV}{\int
\left\langle N_{1},N_{2},t\right\vert \widehat{n}(\mathbf{r}+\mathbf{d}%
/2,t)\left\vert N_{1},N_{2},t\right\rangle \left\langle
N_{1},N_{2},t\right\vert \widehat{n}(\mathbf{r}-\mathbf{d}/2,t)\left\vert
N_{1},N_{2},t\right\rangle dV}=\frac{A}{B}.  \label{correlation_interaction}
\end{equation}%
Here $A$ and $B$ are given in the Appendix.

Based on the expression of $A$ and $B$, we have proven rigorously that $%
g_{nn}\left( \mathbf{d},t\right) \simeq 1$ for $N_{1}\left\vert \zeta
\right\vert \gg 1$ and $N_{2}\left\vert \zeta \right\vert \gg 1$. This flat
behavior for $g_{nn}(\mathbf{d},t)$ can be understood through a simple
analysis. For $N_{1}\left\vert \zeta \right\vert \gg 1$ and $N_{2}\left\vert
\zeta \right\vert \gg 1$, the quantum state (\ref{quantum state sum}) can be
approximated well as%
\begin{equation}
\left\vert N_{1},N_{2}\right\rangle \approx \left\vert N\right\rangle \equiv
\frac{(f^{\dag })^{N_{1}+N_{2}}}{\sqrt{(N_{1}+N_{2})}!}\left\vert
0\right\rangle .  \label{single}
\end{equation}%
Here $f^{\dag }$ is a creation operator which creates a particle in the
single-particle wave function $(\sqrt{N_{1}}\phi _{1}\left( \mathbf{r}%
,t\right) +\sqrt{N_{2}}e^{i\varphi _{c}}\phi _{2}\left( \mathbf{r},t\right)
)/\sqrt{N_{1}+N_{2}}$. This shows the essential physical picture that two
initially independent condensates merge into a single spatially coherent
condensate. For the quantum state given by Eq. (\ref{single}), it is easy to
prove that $g_{nn}(\mathbf{d},t)=1$.

\begin{figure}[tbp]
\includegraphics[width=0.8\linewidth,angle=270]{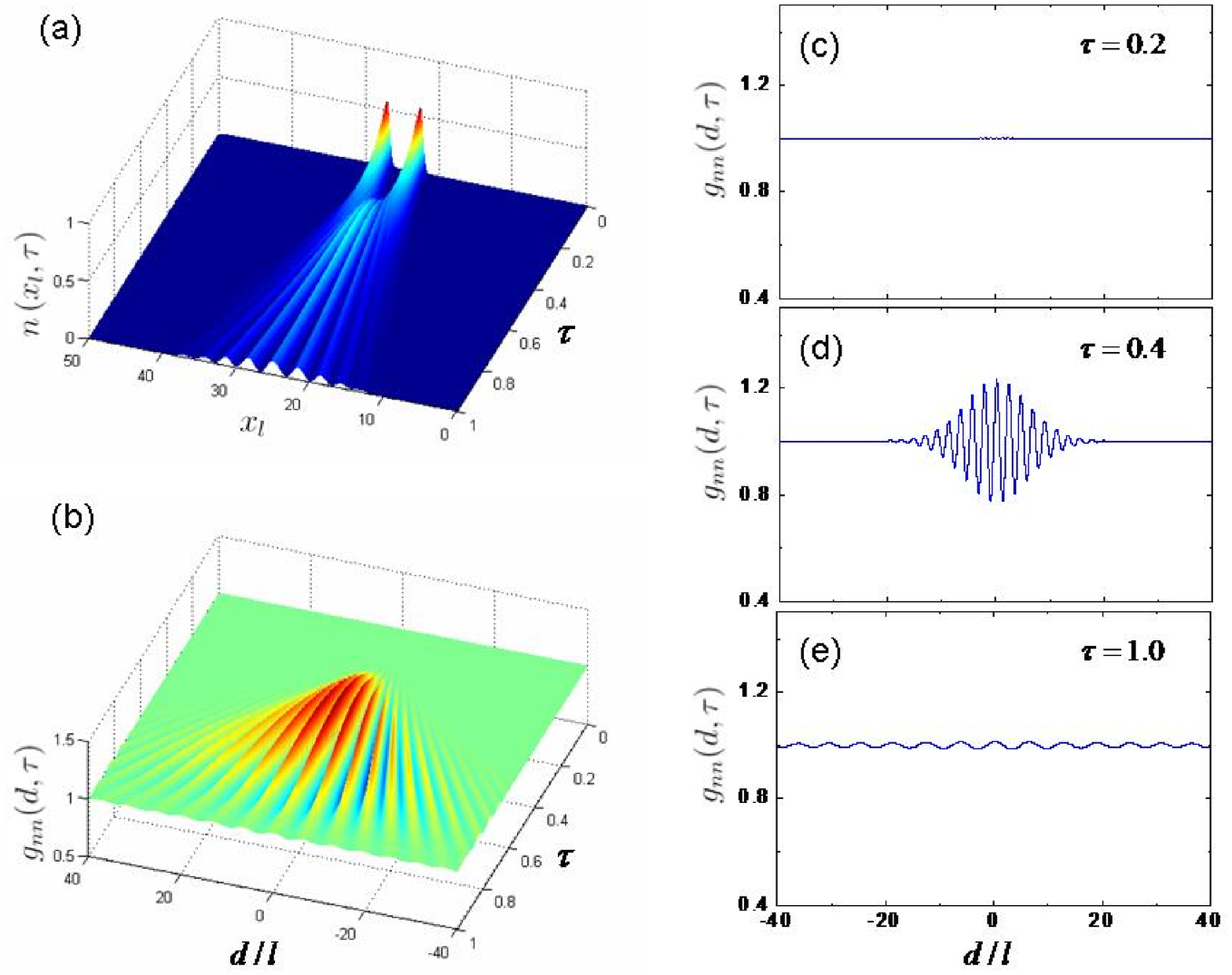}
\caption{(Color online) (a) gives the evolution of the density expectation
value (in unit of $N_{1}+N_{2}$) which shows clear interference patterns
after a full coherence between two condensates is established. (b) gives the
evolution of the density-density correlation function, which shows
interesting feature of emergence and disappearance of the interference
patterns. (c)-(e) show further the density-density correlation function at
different times.}
\end{figure}

We consider the following initial wave functions at $t=0$
\begin{eqnarray}
\phi _{1}\left( x_{l},t=0\right) &=&\frac{1}{\pi ^{1/4}\sqrt{\Delta _{1}}}%
\exp \left[ -\frac{\left( x_{l}-x_{l1}\right) ^{2}}{2\Delta _{1}^{2}}\right]
e^{i\varphi _{1}},  \label{inital1} \\
\phi _{2}\left( x_{l},t=0\right) &=&\frac{1}{\pi ^{1/4}\sqrt{\Delta _{2}}}%
\exp \left[ -\frac{\left( x_{l}-x_{l2}\right) ^{2}}{2\Delta _{2}^{2}}\right]
e^{i\varphi _{2}}.  \label{inital2}
\end{eqnarray}%
In the above wave functions, we have introduced a dimensionless variable $%
x_{l}=x/l$ with $l$ being a length. The factors $\varphi _{1}$ and $\varphi
_{2}$ are two random phases because there is no correlation between two
initially independent Bose condensates. However, one can show that these
random phases play no role in the density distribution and density-density
correlation. We give here an analysis for the case of the density
distribution for brevity, and it is analogous for that of the
density-density correlation. In the interference term (the second term) of
Eq. (\ref{ndensity}), $e^{i\varphi _{c}}=\zeta /\left\vert \zeta \right\vert
$ with $\zeta =\int \phi _{1}\phi _{2}^{\ast }dV$. From the form of the
interference term, it is shown clearly that the terms $e^{i\varphi _{1}}$
and $e^{i\varphi _{2}}$ will be canceled out and play no role in the
interference term of the density distribution.

In the numerical calculations, it is useful to introduce the dimensionless
variable $\tau =E_{l}t/\hslash $ with $E_{l}=\hbar ^{2}/2ml^{2}$, and
dimensionless coupling constants $g_{l1}=N_{1}g/E_{l}l$ and $%
g_{l2}=N_{2}g/E_{l}l$. For the initial wave functions given by Eqs. (\ref%
{inital1}) and (\ref{inital2}), and the parameters $\Delta _{1}=\Delta
_{2}=0.5$, $x_{l2}-x_{l1}=4.5$, $N_{1}=N_{2}=1.0\times 10^{5}$, $%
g_{l1}=g_{l2}=20$, the evolution of $\phi _{1}$ and $\phi _{2}$ is obtained
numerically from Eqs. (\ref{GP1}) and (\ref{GP2}). The evolution of the
density distribution is then given in Fig. 1(a) from the obtained $\phi _{1}$%
, $\phi _{2}$ and Eq. (\ref{ndensity}). With the overlapping of two
condensates, clear interference patterns are shown. Fig. 1(b) gives $%
g_{nn}(d,\tau )$ from obtained $\phi _{1}$, $\phi _{2}$ and Eq. (\ref%
{correlation_interaction}). As shown in Fig. 1(b), interference structure in
$g_{nn}(d,\tau )$ emerges after the initial overlapping between two
condensates, because two condensates are partially coherent. With further
overlapping, however, the interference structure disappears because two
initially independent condensates have merged into a single condensate. This
unique behavior could provide definite signal in an experiment to test the
interaction-induced interference theory. Figs. 1(c)-(e) show further $%
g_{nn}(d,\tau )$ at different times.

Generally speaking, $\left\vert N_{1},N_{2}\right\rangle $ given by Eq. (\ref%
{quantum state sum}) can be regarded as a mixture of coherently superposed
condensate and incoherent condensates. The interference structure in $%
g_{nn}(d,\tau )$ originates from the incoherent component. For the
incoherent component, the interference structure in $g_{nn}(d,\tau )$ due to
incoherent and overlapping condensates is not sensitive to the initial
particle-number fluctuations in two wells. To understand this, one may
recall in the Hanbury-Brown-Twiss stellar interferometer that the light
strength fluctuations of two binary stars in different experiments are not
very important \cite{HBT,Scully}. In addition, the component of coherently
superposed condensate only contributes to a flat behavior in $g_{nn}(d,\tau
) $. Therefore, the interesting behavior shown in Fig. 1(b) is not sensitive
to the initial particle-number fluctuations in two wells.

\section{density-density correlation function in measurement-induced
interference theory}

We now turn to consider the density-density correlation function based on
measurement-induced interference theory. In measurement-induced interference
theory, for the following quantum state of two initially independent
condensates%
\begin{equation}
\left\vert N_{1},N_{2}\right\rangle =\frac{1}{\sqrt{N_{1}!N_{2}!}}\left(
\widehat{a}_{1}^{\dag }\right) ^{N_{1}}\left( \widehat{a}_{2}^{\dag }\right)
^{N_{2}}\left\vert 0\right\rangle ,  \label{Fock state}
\end{equation}%
it is assumed that after the releasing of the double-well potential and even
after the overlapping between two condensates, the quantum state preserves
in this sort of Fock state. In particular, it is assumed that the normalized
wave functions $\phi _{1}$ and $\phi _{2}$ of two condensates are always
orthogonal with each other. For non-interacting situation, this can be
verified rigorously and easily. When $\phi _{1}$ and $\phi _{2}$ are
orthogonal (i.e. $\left[ \widehat{a}_{1},\widehat{a}_{2}^{\dag }\right] =0$%
), it is easy to show that there is no interference term in $\left\langle
N_{1},N_{2}\right\vert \widehat{n}(\mathbf{r},t)\left\vert
N_{1},N_{2}\right\rangle $! Simple calculations give%
\begin{equation}
\left\langle N_{1},N_{2}\right\vert \widehat{n}(\mathbf{r},t)\left\vert
N_{1},N_{2}\right\rangle =N_{1}\left\vert \phi _{1}(\mathbf{r},t)\right\vert
^{2}+N_{2}\left\vert \phi _{2}(\mathbf{r},t)\right\vert ^{2}.
\label{density-Fock}
\end{equation}%
The measurement-induced interference theory tries to find out a physical
mechanism for the observed interference effect missed in the above
expression.

One can prove the following formula%
\begin{equation}
\left\vert N/2,N/2\right\rangle =\left( \frac{\pi N}{2}\right)
^{1/4}\int_{0}^{2\pi }\frac{d\varphi }{2\pi }\left\vert \varphi
,N\right\rangle ,  \label{Fock}
\end{equation}%
where the phase state $\left\vert \varphi ,N\right\rangle $ is given by%
\begin{equation}
\left\vert \varphi ,N\right\rangle =\frac{1}{\left( 2^{N}N!\right) ^{1/2}}%
\left( \widehat{a}_{1}^{\dag }e^{i\varphi /2}+\widehat{a}_{2}^{\dag
}e^{-i\varphi /2}\right) ^{N}\left\vert 0\right\rangle .
\end{equation}%
Because for large particle number, the phase states for different phase
factor $\varphi $ can be approximated well to be orthogonal, it is argued in
the measurement-induced interference theory that in a single-shot
experiment, the result of the density distribution would correspond to that
of a phase state with a random phase factor $\varphi $.

Based on the measurement-induced interference theory, after the overlapping
between two condensates, the measurement makes the quantum state transform
from the Fork state $\left\vert N_{1},N_{2}\right\rangle $ to a fully
coherent phase state $\left\vert \varphi ,N\right\rangle $. In this
situation, the density-density correlation obtained after a series of
experiments takes the form%
\begin{equation}
g_{nn}\left( \mathbf{d},t\right) =\frac{\int_{0}^{2\pi }d\varphi \int
dV\left\langle \varphi ,N\right\vert \widehat{n}(\mathbf{r}+\mathbf{d}/2,t)%
\widehat{n}(\mathbf{r}-\mathbf{d}/2,t)\left\vert \varphi ,N\right\rangle }{%
\int_{0}^{2\pi }d\varphi \int dV\left\langle \varphi ,N\right\vert \widehat{n%
}(\mathbf{r}+\mathbf{d}/2,t)\left\vert \varphi ,N\right\rangle \left\langle
\varphi ,N\right\vert \widehat{n}(\mathbf{r}-\mathbf{d}/2,t)\left\vert
\varphi ,N\right\rangle }.  \label{meas-corr}
\end{equation}%
For $N_{1}>>1$ and $N_{2}>>1$ (which are satisfied for the parameters used
below Eq. (\ref{inital2})), it is easy to find that $g_{nn}\left( \mathbf{d}%
,t\right) \simeq 1$, \textit{i.e.} $g_{nn}\left( \mathbf{d},t\right) $
always shows a flat behavior. It is understandable that this flat behavior
is not sensitive to the initial particle-number fluctuations in $N_{1}$ and $%
N_{2}$.

Combined with Sec. II about the interaction-induced interference theory, we
see that the behavior based on the measurement-induced interference theory
(Eq. (\ref{meas-corr})) is significantly different from the prediction of
the interaction-induced interference theory, where the interference fringes
in the density-density correlation show a behavior of emergence and
disappearance. This suggests an experimental test of the interference
mechanism between the measurement-induced interference theory and
interaction-induced interference theory.

\section{Summary and discussion}

In summary, we consider the density-density correlation for two initially
independent condensates based on both the interaction-induced interference
theory and measurement-induced interference theory. The present work shows
that there is an essential difference in the density-density correlation
between two different theories. In particular, a behavior of emergence and
disappearance for the interference fringes in the density-density
correlation is predicted based on the interaction-induced interference
theory. We suggest the studies of the density-density correlation in future
experiments to test different theories, and thus deep our understanding
about the interference mechanism.

As a last discussion, we consider in brief a recent interesting
experiment about the vortex formation by interference of three
initially independent condensates \cite{vortex}. In this experiment,
three initially independent condensates were first prepared in a
special trapping potential which has a structure of three `petals'.
After the decreasing of the barrier separating three condensates,
the vortex was observed with a probability agreeing well with the
assumption that there are indeterminate relative phases during the
merging of three initially independent condensates. Although the
authors of this experimental paper used the measurement-induced
interference theory to explain the indeterminate relative phases and
merging of three independent condensates into a single condensate,
we think this experiment could be explained more naturally with
interaction-induced quantum merging theory
\cite{xiong1,xiong2,xiong-NJP,xiong-coh}. In this experiment, the
establishment of the relative phase and coherence between different
condensates happen before the measurement of the vortex structure,
i.e. in the whole formation precess of the vertex there is not a
measurement at all. Therefore, it is mysterious to think that a
measurement induces the relative phase and coherence between
initially independent condensates in this experiment about vortex
formation. In the frame of interaction-induced interference
(coherence) theory, the interatomic interaction constructs the
coherence and relative phase between different condensates. Together
the role of damping, a stable vortex can be finally formed. In an
experiment, because different condensates have different particle
number and other indeterminate factors, the relative phase after the
establishment of coherence is random. These analyses show that this
experiment can be explained in the interaction-induced coherence
theory. Further experiments of vortex formation would be quite
interesting if the particle number or interatomic scattering length
are controlled, so that the role of quantum merging due to
interatomic interaction can be tested and studied further.

\begin{acknowledgments}
We acknowledge the useful discussions with B. L. Lu and W. Ketterle. This
work is supported by NSFC under Grant Nos. 10634060, 10474117 and NBRPC
under Grant Nos. 2006CB921406, 2005CB724508 and also funds from Chinese
Academy of Sciences.
\end{acknowledgments}

\newpage

\section*{Appendix}

In this appendix, we give the complex coefficients in $A$ and $B$ for $%
g_{nn}\left( \mathbf{d},t\right) $ given by Eq. (\ref%
{correlation_interaction}). From the form of $g_{nn}\left( \mathbf{d}%
,t\right) $, there are $16$ terms in $A$ and $B$.

\begin{eqnarray}
A &=&\int dV\left[ h_{1}\phi _{1}^{\ast }\left( \mathbf{r}+\mathbf{d}%
/2\right) \phi _{1}\left( \mathbf{r}+\mathbf{d}/2\right) \phi _{1}^{\ast
}\left( \mathbf{r}-\mathbf{d}/2\right) \phi _{1}\left( \mathbf{r}-\mathbf{d}%
/2\right) \right.  \notag \\
&&+h_{2}\phi _{1}^{\ast }\left( \mathbf{r}+\mathbf{d}/2\right) \phi
_{1}\left( \mathbf{r}+\mathbf{d}/2\right) \phi _{1}^{\ast }\left( \mathbf{r}-%
\mathbf{d}/2\right) \phi _{2}^{\prime }\left( \mathbf{r}-\mathbf{d}/2\right)
\notag \\
&&+h_{3}\phi _{1}^{\ast }\left( \mathbf{r}+\mathbf{d}/2\right) \phi
_{1}\left( \mathbf{r}+\mathbf{d}/2\right) \phi _{2}^{\prime \ast }\left(
\mathbf{r}-\mathbf{d}/2\right) \phi _{1}\left( \mathbf{r}-\mathbf{d}/2\right)
\notag \\
&&+h_{4}\phi _{1}^{\ast }\left( \mathbf{r}+\mathbf{d}/2\right) \phi
_{1}\left( \mathbf{r}+\mathbf{d}/2\right) \phi _{2}^{\prime \ast }\left(
\mathbf{r}-\mathbf{d}/2\right) \phi _{2}^{\prime }\left( \mathbf{r}-\mathbf{d%
}/2\right)  \notag \\
&&+h_{5}\phi _{1}^{\ast }\left( \mathbf{r}+\mathbf{d}/2\right) \phi
_{2}^{\prime }\left( \mathbf{r}+\mathbf{d}/2\right) \phi _{1}^{\ast }\left(
\mathbf{r}-\mathbf{d}/2\right) \phi _{1}\left( \mathbf{r}-\mathbf{d}/2\right)
\notag \\
&&+h_{6}\phi _{1}^{\ast }\left( \mathbf{r}+\mathbf{d}/2\right) \phi
_{2}^{\prime }\left( \mathbf{r}+\mathbf{d}/2\right) \phi _{1}^{\ast }\left(
\mathbf{r}-\mathbf{d}/2\right) \phi _{2}^{\prime }\left( \mathbf{r}-\mathbf{d%
}/2\right)  \notag \\
&&+h_{7}\phi _{1}^{\ast }\left( \mathbf{r}+\mathbf{d}/2\right) \phi
_{2}^{\prime }\left( \mathbf{r}+\mathbf{d}/2\right) \phi _{2}^{\prime \ast
}\left( \mathbf{r}-\mathbf{d}/2\right) \phi _{1}\left( \mathbf{r}-\mathbf{d}%
/2\right)  \notag \\
&&+h_{8}\phi _{1}^{\ast }\left( \mathbf{r}+\mathbf{d}/2\right) \phi
_{2}^{\prime }\left( \mathbf{r}+\mathbf{d}/2\right) \phi _{2}^{\prime \ast
}\left( \mathbf{r}-\mathbf{d}/2\right) \phi _{2}^{\prime }\left( \mathbf{r}-%
\mathbf{d}/2\right)  \notag \\
&&+h_{9}\phi _{2}^{\prime \ast }\left( \mathbf{r}+\mathbf{d}/2\right) \phi
_{1}\left( \mathbf{r}+\mathbf{d}/2\right) \phi _{1}^{\ast }\left( \mathbf{r}-%
\mathbf{d}/2\right) \phi _{1}\left( \mathbf{r}-\mathbf{d}/2\right)  \notag \\
&&+h_{10}\phi _{2}^{\prime \ast }\left( \mathbf{r}+\mathbf{d}/2\right) \phi
_{1}\left( \mathbf{r}+\mathbf{d}/2\right) \phi _{1}^{\ast }\left( \mathbf{r}-%
\mathbf{d}/2\right) \phi _{2}^{\prime }\left( \mathbf{r}-\mathbf{d}/2\right)
\notag \\
&&+h_{11}\phi _{2}^{\prime \ast }\left( \mathbf{r}+\mathbf{d}/2\right) \phi
_{1}\left( \mathbf{r}+\mathbf{d}/2\right) \phi _{2}^{\prime \ast }\left(
\mathbf{r}-\mathbf{d}/2\right) \phi _{1}\left( \mathbf{r}-\mathbf{d}/2\right)
\notag \\
&&+h_{12}\phi _{2}^{\prime \ast }\left( \mathbf{r}+\mathbf{d}/2\right) \phi
_{1}\left( \mathbf{r}+\mathbf{d}/2\right) \phi _{2}^{\prime \ast }\left(
\mathbf{r}-\mathbf{d}/2\right) \phi _{2}^{\prime }\left( \mathbf{r}-\mathbf{d%
}/2\right)  \notag \\
&&+h_{13}\phi _{2}^{\prime \ast }\left( \mathbf{r}+\mathbf{d}/2\right) \phi
_{2}^{\prime }\left( \mathbf{r}+\mathbf{d}/2\right) \phi _{1}^{\ast }\left(
\mathbf{r}-\mathbf{d}/2\right) \phi _{1}\left( \mathbf{r}-\mathbf{d}/2\right)
\notag \\
&&+h_{14}\phi _{2}^{\prime \ast }\left( \mathbf{r}+\mathbf{d}/2\right) \phi
_{2}^{\prime }\left( \mathbf{r}+\mathbf{d}/2\right) \phi _{1}^{\ast }\left(
\mathbf{r}-\mathbf{d}/2\right) \phi _{2}^{\prime }\left( \mathbf{r}-\mathbf{d%
}/2\right)  \notag \\
&&+h_{15}\phi _{2}^{\prime \ast }\left( \mathbf{r}+\mathbf{d}/2\right) \phi
_{2}^{\prime }\left( \mathbf{r}+\mathbf{d}/2\right) \phi _{2}^{\prime \ast
}\left( \mathbf{r}-\mathbf{d}/2\right) \phi _{1}\left( \mathbf{r}-\mathbf{d}%
/2\right)  \notag \\
&&\left. +h_{16}\phi _{2}^{\prime \ast }\left( \mathbf{r}+\mathbf{d}%
/2\right) \phi _{2}^{\prime }\left( \mathbf{r}+\mathbf{d}/2\right) \phi
_{2}^{\prime \ast }\left( \mathbf{r}-\mathbf{d}/2\right) \phi _{2}^{\prime
}\left( \mathbf{r}-\mathbf{d}/2\right) \right] .
\end{eqnarray}%
The sixteen coefficients are given by

\begin{eqnarray}
h_{1} &=&\frac{\Xi _{n}^{2}}{N_{1}!N_{2}!\left\vert \beta \right\vert
^{2N_{2}}}\sum\limits_{i=0}^{N_{2}}C_{N_{2}}^{i}C_{N_{2}}^{i}\left(
N_{2}-i\right) !\left( N_{1}+i\right) !\left( N_{1}+i\right) ^{2}\left\vert
\beta \zeta \right\vert ^{2i},  \notag \\
h_{2} &=&\frac{\Xi _{n}^{2}}{N_{1}!N_{2}!\left\vert \beta \right\vert
^{2N_{2}}}\sum\limits_{i=0}^{N_{2}-1}C_{N_{2}}^{i}C_{N_{2}}^{i+1}\left(
N_{2}-i\right) !\left( N_{1}+i+1\right) !\left( N_{1}+i+1\right) \left\vert
\beta \zeta \right\vert ^{2i}\beta ^{\ast }\zeta ,  \notag \\
h_{3} &=&\frac{\Xi _{n}^{2}}{N_{1}!N_{2}!\left\vert \beta \right\vert
^{2N_{2}}}\sum\limits_{i=0}^{N_{2}-1}C_{N_{2}}^{i}C_{N_{2}}^{i+1}\left(
N_{2}-i\right) !\left( N_{1}+i+1\right) !\left( N_{1}+i\right) \left\vert
\beta \zeta \right\vert ^{2i}\beta \zeta ^{\ast },  \notag \\
h_{4} &=&\frac{\Xi _{n}^{2}}{N_{1}!N_{2}!\left\vert \beta \right\vert
^{2N_{2}}}\sum\limits_{i=0}^{N_{2}}C_{N_{2}}^{i}C_{N_{2}}^{i}\left(
N_{2}-i\right) !\left( N_{2}-i\right) \left( N_{1}+i\right) !\left(
N_{1}+i\right) \left\vert \beta \zeta \right\vert ^{2i},  \notag \\
h_{5} &=&\frac{\Xi _{n}^{2}}{N_{1}!N_{2}!\left\vert \beta \right\vert
^{2N_{2}}}\sum\limits_{i=0}^{N_{2}-1}C_{N_{2}}^{i}C_{N_{2}}^{i+1}\left(
N_{2}-i\right) !\left( N_{1}+i+1\right) !\left( N_{1}+i\right) \left\vert
\beta \zeta \right\vert ^{2i}\beta ^{\ast }\zeta ,  \notag \\
h_{6} &=&\frac{\Xi _{n}^{2}}{N_{1}!N_{2}!\left\vert \beta \right\vert
^{2N_{2}}}\sum\limits_{i=0}^{N_{2}-2}C_{N_{2}}^{i}C_{N_{2}}^{i+2}\left(
N_{2}-i\right) !\left( N_{1}+i+2\right) !\left\vert \beta \zeta \right\vert
^{2i}\left( \beta ^{\ast }\zeta \right) ^{2},  \notag \\
h_{7} &=&\frac{\Xi _{n}^{2}}{N_{1}!N_{2}!\left\vert \beta \right\vert
^{2N_{2}}}\sum\limits_{i=0}^{N_{2}}C_{N_{2}}^{i}C_{N_{2}}^{i}\left(
N_{2}-i+1\right) !\left( N_{1}+i\right) !\left( N_{1}+i\right) \left\vert
\beta \zeta \right\vert ^{2i},  \notag \\
h_{8} &=&\frac{\Xi _{n}^{2}}{N_{1}!N_{2}!\left\vert \beta \right\vert
^{2N_{2}}}\sum\limits_{i=0}^{N_{2}-1}C_{N_{2}}^{i}C_{N_{2}}^{i+1}\left(
N_{2}-i\right) !\left( N_{2}-i\right) \left( N_{1}+i+1\right) !\left\vert
\beta \zeta \right\vert ^{2i}\beta ^{\ast }\zeta ,  \notag \\
h_{9} &=&\frac{\Xi _{n}^{2}}{N_{1}!N_{2}!\left\vert \beta \right\vert
^{2N_{2}}}\sum\limits_{i=0}^{N_{2}-1}C_{N_{2}}^{i}C_{N_{2}}^{i+1}\left(
N_{2}-i\right) !\left( N_{1}+i+1\right) !\left( N_{1}+i+1\right) \left\vert
\beta \zeta \right\vert ^{2i}\beta \zeta ^{\ast },  \notag \\
h_{10} &=&\frac{\Xi _{n}^{2}}{N_{1}!N_{2}!\left\vert \beta \right\vert
^{2N_{2}}}\sum\limits_{i=0}^{N_{2}}C_{N_{2}}^{i}C_{N_{2}}^{i}\left(
N_{2}-i\right) !\left( N_{2}-i\right) \left( N_{1}+i+1\right) !\left\vert
\beta \zeta \right\vert ^{2i},  \notag \\
h_{11} &=&\frac{\Xi _{n}^{2}}{N_{1}!N_{2}!\left\vert \beta \right\vert
^{2N_{2}}}\sum\limits_{i=0}^{N_{2}-2}C_{N_{2}}^{i}C_{N_{2}}^{i+2}\left(
N_{2}-i\right) !\left( N_{1}+i+2\right) !\left\vert \beta \zeta \right\vert
^{2i}\left( \beta \zeta ^{\ast }\right) ^{2},  \notag \\
h_{12} &=&\frac{\Xi _{n}^{2}}{N_{1}!N_{2}!\left\vert \beta \right\vert
^{2N_{2}}}\sum\limits_{i=0}^{N_{2}-1}C_{N_{2}}^{i}C_{N_{2}}^{i+1}\left(
N_{2}-i\right) !(N_{2}-i-1)\left( N_{1}+i+1\right) !\left\vert \beta \zeta
\right\vert ^{2i}\beta \zeta ^{\ast },  \notag \\
h_{13} &=&\frac{\Xi _{n}^{2}}{N_{1}!N_{2}!\left\vert \beta \right\vert
^{2N_{2}}}\sum\limits_{i=0}^{N_{2}}C_{N_{2}}^{i}C_{N_{2}}^{i}\left(
N_{2}-i\right) !\left( N_{2}-i\right) \left( N_{1}+i\right) !\left(
N_{1}+i\right) \left\vert \beta \zeta \right\vert ^{2i},  \notag \\
h_{14} &=&\frac{\Xi _{n}^{2}}{N_{1}!N_{2}!\left\vert \beta \right\vert
^{2N_{2}}}\sum\limits_{i=0}^{N_{2}-1}C_{N_{2}}^{i}C_{N_{2}}^{i+1}\left(
N_{2}-i\right) !\left( N_{2}-i-1\right) \left( N_{1}+i+1\right) !\left\vert
\beta \zeta \right\vert ^{2i}\beta ^{\ast }\zeta ,  \notag \\
h_{15} &=&\frac{\Xi _{n}^{2}}{N_{1}!N_{2}!\left\vert \beta \right\vert
^{2N_{2}}}\sum\limits_{i=0}^{N_{2}-1}C_{N_{2}}^{i}C_{N_{2}}^{i+1}\left(
N_{2}-i\right) !\left( N_{2}-i\right) \left( N_{1}+i+1\right) !\left\vert
\beta \zeta \right\vert ^{2i}\beta \zeta ^{\ast },  \notag \\
h_{16} &=&\frac{\Xi _{n}^{2}}{N_{1}!N_{2}!\left\vert \beta \right\vert
^{2N_{2}}}\sum\limits_{i=0}^{N_{2}}C_{N_{2}}^{i}C_{N_{2}}^{i}\left(
N_{2}-i\right) !\left( N_{2}-i\right) ^{2}\left( N_{1}+i\right) !\left\vert
\beta \zeta \right\vert ^{2i}.
\end{eqnarray}

$B$ also comprises sixteen coefficients. $B$ can he obtained by replacing $%
h_{i}$ in $A$ by the following $p_{i}$ (i=1,$\cdot \cdot \cdot $,16):

\begin{eqnarray}
p_{1}
&=&a^{2},p_{2}=ab,p_{3}=ac,p_{4}=ad,p_{5}=ab,p_{6}=b^{2},p_{7}=bc,p_{8}=bd,
\notag \\
p_{9}
&=&ac,p_{10}=bc,p_{11}=c^{2},p_{12}=cd,p_{13}=ad,p_{14}=bd,p_{15}=cd,p_{16}=d^{2}.
\end{eqnarray}%
Here%
\begin{eqnarray}
a &=&\frac{\Xi _{n}^{2}}{N_{1}!N_{2}!\left\vert \beta \right\vert ^{2N_{2}}}%
\sum\limits_{i=0}^{N_{2}}C_{N_{2}}^{i}C_{N_{2}}^{i}\left( N_{2}-i\right)
!\left( N_{1}+i\right) !\left( N_{1}+i\right) \left\vert \beta \zeta
\right\vert ^{2i},  \notag \\
b &=&\frac{\Xi _{n}^{2}}{N_{1}!N_{2}!\left\vert \beta \right\vert ^{2N_{2}}}%
\sum\limits_{i=0}^{N_{2}-1}C_{N_{2}}^{i}C_{N_{2}}^{i+1}\left( N_{2}-i\right)
!\left( N_{1}+i+1\right) !\left\vert \beta \zeta \right\vert ^{2i}\beta
^{\ast }\zeta ,  \notag \\
c &=&\frac{\Xi _{n}^{2}}{N_{1}!N_{2}!\left\vert \beta \right\vert ^{2N_{2}}}%
\sum\limits_{i=0}^{N_{2}-1}C_{N_{2}}^{i}C_{N_{2}}^{i+1}\left( N_{2}-i\right)
!\left( N_{1}+i+1\right) !\left\vert \beta \zeta \right\vert ^{2i}\beta
\zeta ^{\ast },  \notag \\
d &=&\frac{\Xi _{n}^{2}}{N_{1}!N_{2}!\left\vert \beta \right\vert ^{2N_{2}}}%
\sum\limits_{i=0}^{N_{2}}C_{N_{2}}^{i}C_{N_{2}}^{i}\left( N_{2}-i\right)
!\left( N_{2}-i\right) \left( N_{1}+i\right) !\left\vert \beta \zeta
\right\vert ^{2i}.
\end{eqnarray}

For $N_{1}\left\vert \zeta \right\vert \gg 1$ and $N_{2}\left\vert \zeta
\right\vert \gg 1$, we have verified by numerical calculations that $%
h_{i}\simeq p_{i}$ (i=1,$\cdot \cdot \cdot $,16). Therefore, we have $%
g_{nn}\left( \mathbf{d},t\right) \simeq 1$ in this situation.


\begin{thebibliography}{99}
\bibitem{note1} Based on the interaction-induced interference theory, after
two initially independent condensates merge into a single condensate, the
order parameter can be approximated as $\Psi \left( \mathbf{r},t\right)
=\left( \sqrt{N_{1}}\phi _{1}e^{-i\varphi }+\sqrt{N_{2}}\phi _{2}\right) $.
Here $\varphi $ is the relative phase after the quantum merging process.
After full quantum merging, we have $d\varphi /dt=\Delta \mu t/\hbar $ with $%
\Delta \mu $ being the difference of the chemical potential of two
condensates. Thus, after full quantum merging, we have $\varphi \left(
t\right) =\Delta \mu t/\hbar +\varphi _{0}$. Here $\varphi _{0}$ is a phase
arising from the quantum merging process. By using the experimental
parameters in \cite{Andrew} and $\mu =15^{2/5}\left( Na/a_{ho}\right)
^{2/5}\hbar \omega _{ho}/2$ (Here $\omega _{ho}=\left( \omega _{x}\omega
_{y}\omega _{z}\right) ^{1/3}$. $a$ is the scattering length and $a_{ho}=%
\sqrt{\hbar /m\omega _{ho}}$.), it is easy to get $\left\vert \Delta \mu
t/\hbar \right\vert \approx 20\pi \left\vert \triangle N/N\right\vert $.
\bibitem{note2} In \cite{Andrew}, the initial particle number before the
evaporative cooling is about $5\times 10^{9}$ (see also \cite%
{condensate-ketterle}). After the evaporative cooling, one excepts the
particle-number fluctuations in the condensate are at least Poissonian \cite%
{note3}. Thus, we have $\left\vert \Delta N\right\vert >7\times 10^{4}$. For
the particle number of about $2.5\times 10^{6}$ in each condensate, we have $%
\left\vert \triangle N/N\right\vert >2\%$.
\end{thebibliography}
\end{document}